# Baseline-free quantitative absorption spectroscopy based on cepstral analysis

RYAN K. COLE[†], AMANDA S. MAKOWIECKI[†], NAZANIN HOGHOOGHI, AND GREGORY B. RIEKER*

Precision Laser Diagnostics Laboratory, Department of Mechanical Engineering, University of Colorado Boulder, Boulder, CO 80309, USA
*greg.rieker@colorado.edu
[†]Both authors contributed equally to this work

**Abstract:** The accuracy of quantitative absorption spectroscopy depends on correctly distinguishing molecular absorption signatures in a measured transmission spectrum from the varying intensity or 'baseline' of the light source. Baseline correction becomes particularly difficult when the measurement involves complex, broadly absorbing molecules or non-ideal transmission effects such as etalons. We demonstrate a technique that eliminates the need to account for the laser intensity in absorption spectroscopy by converting the measured transmission spectrum of a gas sample to a modified form of the time-domain molecular free induction decay (m-FID) using a cepstral analysis approach developed for audio signal processing. Much of the m-FID signal is temporally separated from and independent of the source intensity, and this portion can be fit directly with a model to determine sample gas properties without correcting for the light source intensity. We validate the new approach in several complex absorption spectroscopy scenarios and discuss its limitations. The technique is applicable to spectra obtained with any absorption spectrometer and provides a fast and accurate approach for analyzing complex spectra.

## 1. Introduction

Absorption spectroscopy is an important technique for quantitative, nonintrusive measurement in a variety of systems. Molecular absorption spectroscopy measures the amount of light absorbed at frequencies resonant with molecular quantum state transitions. In sensing applications, the magnitude and shape of the measured absorption signatures can be fit with a model of the expected absorption as a function of the environmental conditions (temperature, pressure, and absorber concentration) in order to obtain a measurement of the conditions in the sample gas.

Quantitative comparison of a measured transmission spectrum to an absorption model requires that the measured spectrum be normalized by the non-absorbing intensity spectrum of the light source (the "baseline" intensity). Though not often discussed in the literature, this normalization process can be the limiting factor in the ultimate accuracy and stability of absorption-based measurements. Accurate normalization of the baseline is complicated by variations in the light source intensity with wavelength and time as well as changes in the system transmission due to scatter, beam-steering, and window interference (e.g. etalons). Accurate normalization becomes even more difficult when measuring broadly absorbing molecular species or when inaccuracy exists in the spectral absorption models used for fitting the measured absorption features.

Many baseline-insensitive approaches to absorption spectroscopy have been developed, as described in the next section. These approaches require specific hardware or optical setups, and are generally applicable in specialized situations (e.g. where sensing can be implemented with optical cavities or paramagnetic molecules, etc.). Here, we demonstrate a baseline-insensitive approach that can be used with any absorption spectrometer. We adapt a cepstral analysis technique that has been used in speech and audio signal processing for decades [1–3] to convert the measured transmission spectrum of a gas sample to a modified form of the time-domain free induction decay signal, which we refer to as the modified free induction decay (m-FID).

In this approach, we develop a signal in which much of the molecular response is both temporally separated from and independent of the source intensity.

We demonstrate the m-FID technique in several challenging absorption spectroscopy scenarios. First, we apply the technique to accurately determine absorber concentration in a dual frequency comb measurement of a spectrally overlapping and broadly absorbing mixture of ethane and methane. We then extend these results to measure species concentrations from a simulated, broadband, mid-infrared spectrum of a mixture of four broadly absorbing molecular species, including significant etalon interference and a highly non-uniform source intensity. We show that when the spectrum is degraded with a complex baseline and intensity noise, the technique greatly improves the measurement accuracy compared to a traditional, frequency-domain fitting approach.

In all of these scenarios, we significantly reduce the degrees of freedom of the fit compared to a frequency-domain routine. This results in an increase in fit speed (up to a factor of twenty for the datasets shown in this paper), and removes the uncertainty and inaccuracy associated with these degrees of freedom and the user judgement involved in selecting baseline fitting parameters. While the measurements presented in this paper were made using a dual frequency comb spectrometer, the technique is applicable for any absorption spectrometer because the m-FID signal can be constructed from any measured transmission spectrum.

## 2. Existing methods for baseline normalization

Several successful laser-based techniques have been developed that are insensitive to the baseline intensity of the light source, such as cavity ring-down spectroscopy [4], chirped laser dispersion spectroscopy [5], and Faraday rotation spectroscopy [6]. These techniques circumvent the need for baseline correction by measuring quantities other than the attenuation of light, such as decay time, dispersion, or polarization. These techniques have enabled accurate and precise sensing in many applications, but have specific requirements that favor certain situations. For example, cavity ring down and chirped laser dispersion spectroscopy are challenging to implement in a resolved, broadband configuration, and Faraday rotation spectroscopy is sensitive only to paramagnetic molecules.

Other spectroscopic techniques such as polarization [7] and photoacoustic spectroscopy [8] or laser-induced fluorescence [9] are 'background free' (meaning the measured signal is zero in the absence of a molecular sample), however the measured signal still requires normalization by the laser source intensity to deliver absolute measurements. Derivative techniques of absorption spectroscopy such as wavelength and frequency modulation spectroscopy are background free and provide a means of normalizing the laser source intensity across different modulation harmonics [10–12], though they require a laser source with rapid wavelength tuning capability.

For direct absorption spectroscopy, there are three primary approaches to correct for the light source intensity spectrum: measuring a non-absorbing reference transmission spectrum to later normalize the transmission spectrum of the sample, fitting the baseline in post-processing, or using time-domain techniques to experimentally cancel the signal corresponding to the baseline intensity spectrum.

The reference spectrum approach involves directly measuring the non-absorbing intensity spectrum of the light source with the target sample removed. In practice, this technique is rarely used alone (i.e. without additional baseline correction) because very minor changes to the optical path can alter the transmitted intensity spectrum. It is also often infeasible in many practical situations, for example in large combustion devices, open-path atmospheric measurements, or if there is temporal drift in the intensity spectrum of the source. As such, many spectrometers instead require correction for the background intensity in post-processing.

The most common post-processing techniques involve fitting a polynomial or spline to the 'non-absorbing' parts of the measured spectrum or by masking molecular absorption features with a known spectral model during the baseline fit [13–20]. While these approaches are effective in many applications, they have several important drawbacks. Fitting a function to the non-absorbing parts of the intensity spectrum requires user input/decision-making and comes with associated error/uncertainty. Furthermore, in practice the intensity spectrum that must be modeled can be very complex due to intensity modulation caused by non-linear broadening processes (in broadband lasers) or etalon interference effects introduced through planar optical components (e.g. windows) common in real-world systems. Depending on the degree of variation in the intensity profile, fitting a function to a complex baseline can dramatically increase computation time by requiring high-order polynomials or a large number of individual segments that increase the degrees of freedom that must be optimized [14]. The use of complicated baseline functions can lead to erroneous effects in the baseline-corrected spectrum due to coupling between the modeled baseline and the spectral model used to mask absorption features (e.g. forcing the data to match an incorrect spectral model by changing the modeled baseline). Most importantly, polynomial fitting techniques do not perform well for broadly absorbing molecules (e.g. high-pressure gases or large hydrocarbons), where broadband absorption features force the fitting routine to interpolate the baseline for tens to hundreds of wavenumbers without being informed by non-absorbing spectral regions.

Other post-processing techniques correct for the baseline by relying on shape differences between the non-absorbing intensity spectrum of the light source and the molecular absorption features. For instance, a clever approach by Kranendonk *et al.* [21] reduces sensitivity to the light source intensity spectrum by analyzing the first derivative of the absorption spectrum. Because the variations in the light source intensity spectrum are typically less sharp compared to molecular absorption features (for small molecules), the differentiation process suppresses the baseline contribution. However, this method requires spectral smoothing to compensate for the high frequency noise added in differentiation. As a second example, the Fourier Transform Method for baseline correction [19,22,23] takes a Fourier transform of the measured transmission spectrum, bandpass filters the result to isolate the absorption features (which appear at intermediate frequencies for small molecules), and then uses an inverse Fourier transform to yield a baseline-corrected spectrum. The approach does not remove the dependence of the molecular response on the laser intensity, so the measured absorption features are still scaled by the variable baseline intensity and thus cannot be quantitatively compared to an absorption model without prior knowledge or fitting of the baseline.

An alternative to traditional absorption spectroscopy is time-domain spectroscopy, which measures the free induction decay of molecules after excitation by a pulse of radiation [24–26]. In the infrared, the free induction decay can be directly measured by dual-comb spectroscopy [27,28] or using time-delayed pulses and mixing techniques [29–31]. Because much of the molecular response is temporally separated from the laser excitation, the approach shows promise as a background-free technique. However, the magnitude of the free induction decay signal is still related to the intensity of the excitation pulse [30] and therefore must be accounted for in a quantitative fit to extract gas properties. Recently, Tomberg *et al.* [32] demonstrated a baseline-free technique for broadband absorption spectroscopy using a Michelson interferometer. In this technique, light from the spectrometer that has passed through a gas sample is interfered with light from a reference channel that has been sign-inverted [33–37]. Interference of the two signals effectively cancels the central peak of the interferogram (containing the background intensity spectrum) while leaving the free induction decay. This approach requires careful matching of the optical path delay on the experimental and reference channels, which may inhibit its application in noisy or turbulent environments.

### 3.   Modified free induction decay

The baseline-free approach described in this paper can be viewed in terms of the time-domain description of absorption spectroscopy. In the time-domain, a short laser pulse resonant with the ro-vibrational states of a molecular sample excites an ensemble of molecules that behave as vibrating, rotating dipoles [24–26,29,38–40]. In this process, the transfer of energy via photon absorption from the driving pulse into the molecules at specific frequencies leads to the characteristic absorption signatures in the transmitted laser spectrum [30]. In the time-domain, because all of the molecules are excited at the same instant, the ensemble of molecules initially rotate in phase and radiate via dipole emission into the same mode as the laser source. Because the broadband pulse excites the molecules to a number of different states, the excited molecules rotate at different rates fixed by their rotational quanta. The differing rotation rates cause the molecules to quickly rotate out phase with each other, and thus no longer coherently emit into the laser mode. However, because the rotation rates are quantized, the excited states re-phase a short time later, leading to another coherent forward pulse of radiation [25,38–40]. Depending on the molecular ro-vibrational structure and collision dynamics, the coherent emission from the excited molecules ("echoes" [25,39]) can persist for several nanoseconds after the initial laser pulse. This pattern of dipole emission decays exponentially following the laser pulse, and has been referred to as optical free induction decay (FID) [26,29,39].

An illustration of a simple optical FID signal is shown in Fig. 1. In the time-domain, the electric field corresponding to each excited state is modeled as a decaying waveform oscillating at the resonant frequency of the transition. Examples of these decaying waveforms are shown as gray traces in Fig. 1. For a large number of excited transitions, the decaying waveforms constructively interfere at certain intervals, which in the molecular picture is equivalent to the excited molecules re-phasing to form a pulse of coherent forward scattered radiation (see Fig. 1). As such, the complete optical free induction decay signal assuming optically thin conditions and an instantaneous pulse of exciting radiation with unit intensity can be modeled as [30,39]

$$I(t,L,P) = 1 - \rho L \Delta t \sum_i 2S_i\, \Theta(t) \cos(2\pi \nu_i t) \exp\left(-\frac{\pi^2 \gamma_{D,i}^2 t^2}{\ln 2} - 2\pi \gamma_{L,i} P t\right) \quad (1)$$

where $\rho$ is the number density, $L$ is the laser path length through the sample, $\Delta t$ is the time-domain point spacing, $S$ is the linestrength, $\Theta$ is the Heaviside function, $\nu$ is the quantum transition frequency, $\gamma_D$ is the Doppler half-width, $\gamma_L$ is the total Lorentzian half-width, $P$ is the pressure, and the sum is over all transitions $i$. Eq. (1) assumes the linestrength and broadening terms have units and temperature dependences consistent with the HITRAN database [41,42]. If the optically thin conditions required to implement Eq. (1) are not met, the free induction decay can instead be modeled as the inverse Fourier transform of the transmission spectrum of the gas sample [30,38], which can be created from published absorption databases.

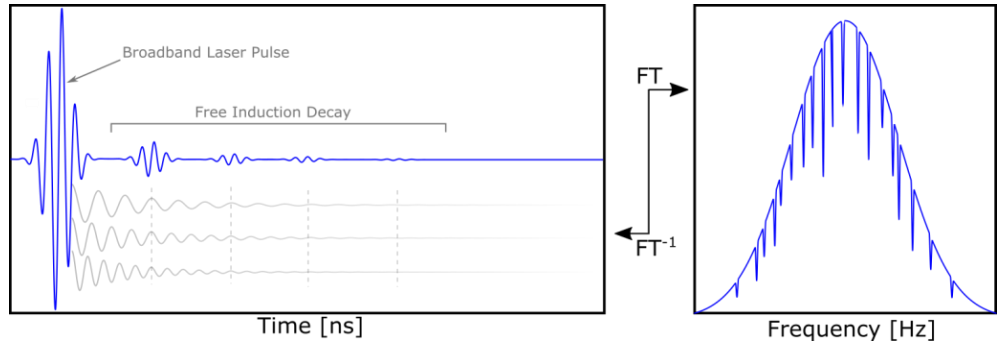

Fig. 1. (Left panel) An example molecular free induction decay signal. The ensemble of excited molecules coherently radiate via dipole emission into the laser mode, leading to a weak, decaying pulse train of emitted light following the incident laser pulse. Each excited state can be modeled as a decaying waveform with a frequency equal to the transition frequency (shown as gray traces in the figure). In this picture, the waveforms periodically interfere at certain intervals to form coherent pulses of light from the excited molecules. (Right panel) The molecular free induction decay signal is related to the traditional transmission spectrum of the gas sample via an inverse Fourier transform.

While much of the free induction decay is, by nature, temporally separated from the excitation laser pulse, it is not fully independent of the source intensity. The free induction decay signal can be inferred from the inverse Fourier transform of the transmission spectrum, which is described by Beer's law (Eq. (2)). Eq. (3) shows that when forming the free induction decay signal from the transmission spectrum, the source intensity is convolved with the molecular absorption response:

$$I(v) = I_o(v)e^{-\alpha(v)} \qquad (2)$$

$$I(t) = F^{-1}[I(v)] = F^{-1}[I_o(v)] * F^{-1}[e^{-\alpha(v)}] \qquad (3)$$

where $I(v)$ is the measured transmission spectrum, $I(t)$ is the corresponding time-domain free induction decay signal, $I_o(v)$ is the baseline source intensity, F denotes the Fourier transform, $*$ denotes convolution, and $\alpha(v) = k_v L$ is absorbance with absorption coefficient $k_v$. The convolution of the molecular absorption response ($F^{-1}[e^{-\alpha(v)}]$) with the baseline source intensity signal ($F^{-1}[I_o(v)]$) makes the two components difficult to separate because the intensity signal affects every point in the free induction decay. This effect is illustrated in Fig.

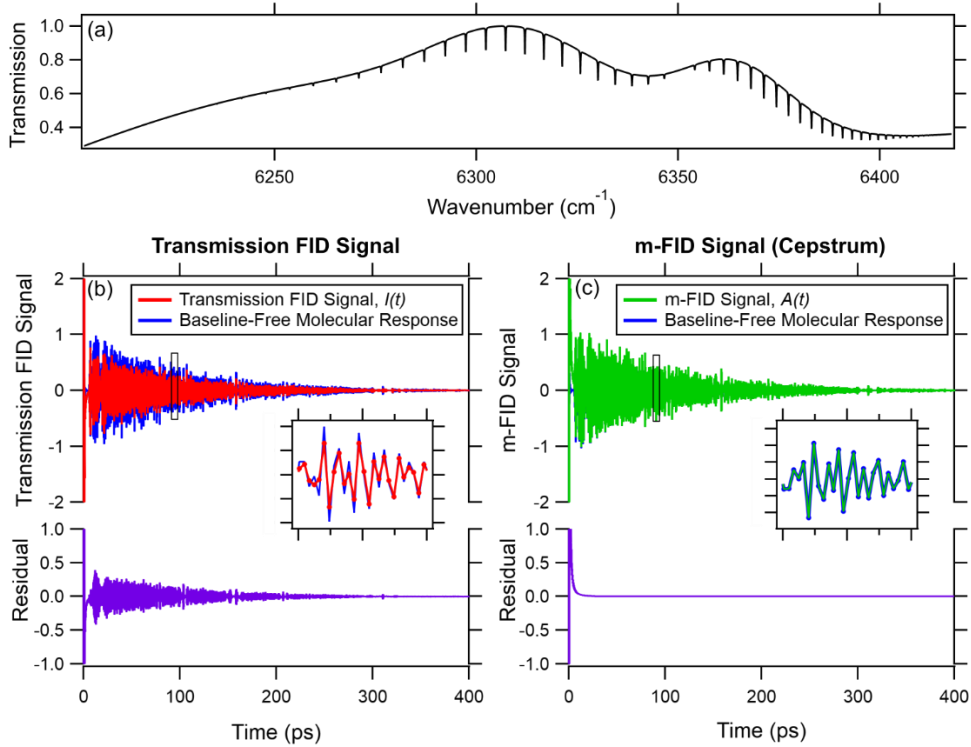

Fig. 2. Comparison of the traditional time-domain FID signal and the m-FID signal generated from the same simulated spectrum of carbon monoxide. (a) The carbon monoxide transmission spectrum simulated at atmospheric conditions with an arbitrary baseline light source intensity profile. (b) The carbon monoxide FID signal derived as the inverse Fourier transform of the spectrum in panel (a). The significant differences between the FID signal and the baseline-free molecular response, shown in the residual plot at the bottom of (b), indicate that the FID signal is still influenced by the baseline intensity. (c) The m-FID signal (cepstrum) generated from the spectrum in (a). The excellent correspondence between the m-FID signal and the baseline-free molecular response over the majority of the signal indicates that the molecular response is largely separated from and independent of the baseline intensity in the m-FID signal. The baseline light source intensity contribution is the rapidly decaying portion in the first few ps of the residual.

2(b) for the simulated free induction decay signal of carbon monoxide calculated as the inverse Fourier transform of the simulated transmission spectrum of Fig. 2(a). In this case, the simulated intensity profile in Fig. 2(a) causes significant differences between the FID signal (which is influenced by the baseline intensity) and the baseline-free molecular response. For this reason, the source intensity spectrum would still need to be explicitly accounted for in a fit of the FID signal.

If instead one takes the negative natural logarithm of the transmission spectrum before applying the inverse Fourier transform, an additive relationship forms between the molecular response and the source intensity spectrum (Eq. (4)). The inverse Fourier Transform maintains this additive relationship and eliminates the convolution in the traditional time-domain FID signal (Eq. (5)):

$$A(\nu) = -\ln(I(\nu)) = \alpha(\nu) - \ln(I_o(\nu)) \tag{4}$$

$$A(t) = \mathcal{F}^{-1}[-\ln(I(\nu))] = \mathcal{F}^{-1}[\alpha(\nu)] + \mathcal{F}^{-1}[-\ln(I_o(\nu))] \tag{5}$$

In Eq. (5), the measured time-domain signal $A(t)$ is what we call the "modified free induction decay" (m-FID), which is now the *addition* of the molecular absorption response, $\mathcal{F}^{-1}[\alpha(\nu)]$, with the source intensity signal. In the language of signal processing, $A(t)$ is the cepstrum of the time-domain transmission signal, $I(t)$ [1,3]. In practice, the component of the signal corresponding to the source intensity decays to zero very rapidly so that the measured signal at longer times is dominated by the molecular response. This effect is shown in Fig. 2(c) for the simulated m-FID signal of carbon monoxide generated from the simulated spectrum of Fig. 2(a). Here, there is excellent correspondence between the m-FID signal (which includes baseline effects) and the baseline-free molecular response over the majority of the m-FID signal. The baseline intensity component of the m-FID signal (the rapidly decaying portion at early times in Fig. 2(c)) is confined to the earliest parts of the signal and is deconvolved from the molecular response. For this reason, much of the m-FID signal is baseline-free, meaning the molecular response is separated from and independent of the baseline intensity.

## 4. Modified free induction decay analysis

The cepstral analysis approach described in the preceding section results in a signal (the m-FID) in which the influence of the baseline is deconvolved and largely separated in time from the majority of the molecular response. In quantitative sensing applications, this portion of the m-FID signal that is uninfluenced by the baseline intensity can be fit with a known, condition-specific model to extract the temperature, pressure, or absorber concentration without any knowledge of, or correction for, the baseline source intensity.

The process to construct and fit an m-FID signal is outlined in Fig. 3, and begins by taking the negative natural logarithm of the measured transmission spectrum of the gas sample. Transmission spectra with non-uniform point spacing must first be re-sampled to a uniform spacing. An inverse Fourier transform then yields the cepstrum, which is made up of both the molecular response and the laser intensity signals. A Levenberg-Marquardt algorithm is used to fit a time-domain model generated from the HITRAN database [41] via the HAPI interface [44] to the m-FID signal. The model is first generated as an absorbance spectrum at an estimated temperature, pressure, and composition, and then converted to the time-domain free induction decay via an inverse Fourier transform. If the conditions are optically thin, the free induction decay model can be generated directly using Eq (1). In order to fit only the molecular response, a weighting function is used to exclude the early portion of the signal that is influenced by the baseline intensity. Here, we choose the simple weighting function

$$W(t) = \begin{cases} 0 & t < t_1 \\ 1 & t_1 \leq t \leq t_2 \\ 0 & t > t_2 \end{cases} \tag{6}$$

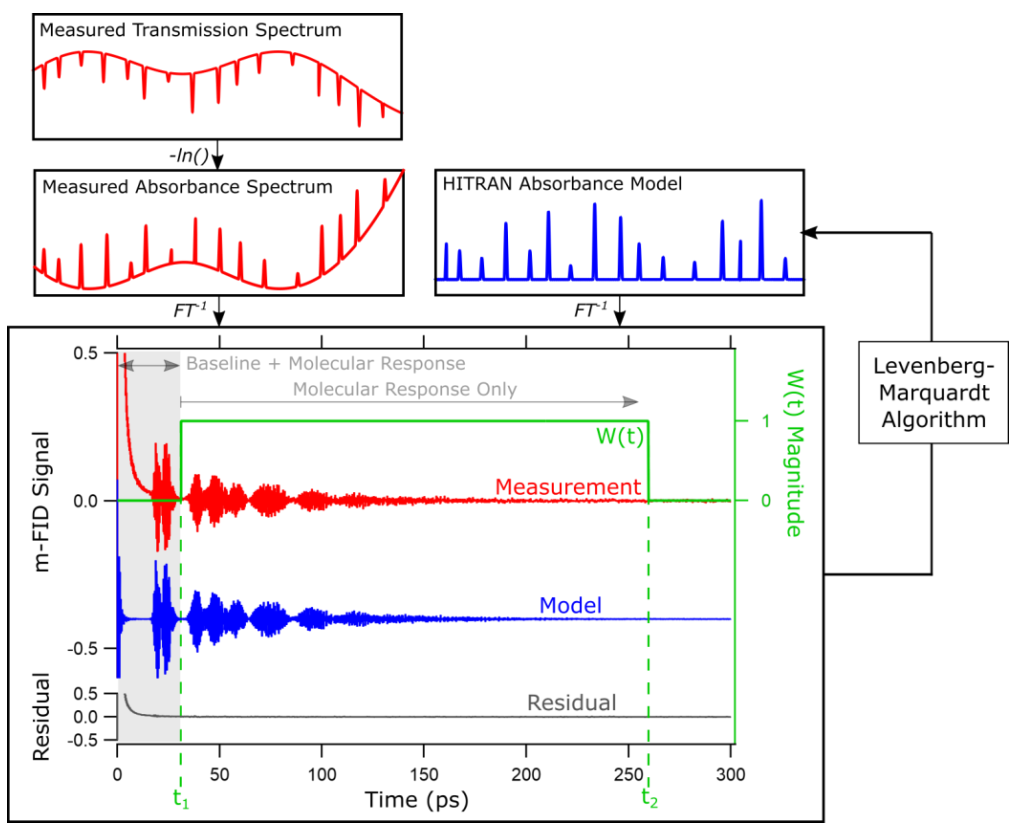

Fig. 3: Schematic of the m-FID fitting approach. The modified free induction decay signal is generated through an inverse Fourier transform of a measured, non-normalized absorbance spectrum. The HITRAN-based model (offset for clarity) is fit to the measured m-FID through a Levenberg-Marquardt algorithm. The fitting algorithm excludes the contribution from the background laser intensity (the far left of the measured signal) by using a weighing function W(t).

The parameter $t_1$ is set such that the fit does not consider the baseline source intensity contribution to the signal, which only significantly overlaps with the molecular response at the earliest times in the measured signal. In practice, an appropriate value of $t_1$ can be estimated from a background dataset (which need not be a perfect representation of the transmitted baseline intensity) or can be informed by the time at which the mean measured m-FID signal deviates from zero (indicative of significant baseline contribution). $t_2$ is an optional parameter, and can be selected as the point at which the molecular response decays to the noise level of the m-FID signal. When the $t_2$ parameter is used, the fitting approach is not sensitive to noise beyond the weighting window, which corresponds to high frequency noise in the transmission spectrum. Fig. 3 shows typical choices for $t_1$ and $t_2$ relative to a simulated m-FID signal. In the time-domain, etalon interference effects manifest as a sharp signal at a fixed time given by the round trip light propagation delay between the optical components generating the interference (see Fig. 6 for an example). In this case, additional 'windows' can be added to the weighting function to gate out the non-ideal portion of the signal corresponding to the etalon, thus keeping these effects from being considered by the fitting algorithm.

By excluding the early portion of the m-FID signal (i.e. the portion that is influenced by the baseline intensity), the weighting function also necessarily excludes a small portion of the molecular response from the fitting algorithm. This information loss leads to a slight degradation to the m-FID concentration detection limit, which is assessed in Appendix A. Similarly, the m-FID analysis technique requires conditions in which the molecular response signal persists for longer than the source intensity signal. In practice, this requirement is not satisfied when the measurement involves molecules without any resolved structure, such as liquids or extremely high pressure gases. For example, as density is increased, molecular collisions become more frequent. These collisions dephase the ensemble of vibrating, rotating molecules, which shortens the time over which the FID signal is present. In the context of the m-FID approach, shortening the time over which the FID signal is present and separated from the intensity signal reduces the number of points that are available for fitting. In these cases, it becomes increasingly important to minimize modulation in the source intensity spectrum, thereby shortening the source intensity component in the cepstrum. This minimizes the amount of time that the intensity signal overlaps with the free induction decay and maximizes the number of points available for fitting.

## 5. Validation: Concentration measurement in a mixture of methane and ethane

We test the m-FID approach by applying the technique to measure the composition of an ethane-methane mixture, which absorbs continuously for more than 500 cm$^{-1}$ in the near-infrared. Accurately measuring concentration from this spectrum (and the spectra of other broadly absorbing gases) is challenging using existing analysis techniques. Because of the broadband absorption, a post-processing routine to fit the baseline of the ethane-methane spectrum would need to interpolate for more than 500 cm$^{-1}$. This makes the modeled baseline particularly sensitive to errors in the spectral model, since a traditional frequency-domain fit to this mixture would almost certainly require the baseline and absorption model to be fit to the data simultaneously. Furthermore, the lack of non-absorbing regions makes it very difficult to identify baseline errors from unintended drift in the laser intensity spectrum or from poorly selected baseline model parameters (e.g. polynomial coefficients). Regardless of the cause, errors in the baseline used to normalize the measured spectrum result directly in errors in the measured absorber concentrations. Because the m-FID approach does not require any baseline correction, the technique is particularly well-suited to applications involving broadly absorbing molecules.

The ethane-methane mixture used for this test is composed of $3.8 \pm 1$% methane in a balance of $96.2 \pm 1.1$% ethane. The uncertainty in the known mole fractions is primarily driven by the uncertainty of the pressure transducer used to create the mixture (0.5% of reading). We create the mixture by filling a stainless steel mixing tank while monitoring the pressure using a calibrated capacitance manometer (MKS Baratron 722B). Before filling the $453 \pm 4$ mm quartz optical cell, we agitate the mixture with stainless steel ball bearings in the mixing tank to ensure uniformity. We measure the transmission spectrum of the mixture at $297.5 \pm 2.2$ K and $630.8 \pm 3.2$ Torr pressure using a near-infrared dual comb spectrometer with 0.0066 cm$^{-1}$ point spacing and coherent averaging for 90 minutes. The pressure and temperature of the mixture remain stable for the duration of the measurement.

We supply the m-FID fitting algorithm with an initial guess for the mixture mole fraction of 2% methane, while the temperature and pressure are fixed at the known experimental values. The details of the absorption model used to fit the mixture spectrum are given in Appendix B.

Figure 4 shows the fit to the measured m-FID signal in the time-domain (panel (a)). Panel (b) shows the fit results in the frequency-domain, which are obtained through the Fourier transform of the time-domain traces in panel (a). In Fig. 4(b), the frequency-domain view of the fit shows how the m-FID approach is able to separate the molecular response from the background laser intensity spectrum. Here, the 'Fit' trace is the Fourier transform of the m-FID model (which includes no baseline or laser intensity parameters) that has been fit to the measured time-domain signal. The 'Residual' trace is composed of the true source intensity baseline along with any differences between the measured and modeled absorption features. As such, Fig. 4(b) shows an interesting side effect of the fact that the m-FID approach does not require or take into account the baseline laser intensity during the fitting process. The fit residual is an interesting (and potentially useful) way of measuring the source intensity baseline or assessing model-data mismatch. The latter includes both the ability to assess the accuracy of absorption models or to determine whether one is accounting for all of the molecules that are present in the measurement (e.g. the absorption spectrum of any molecules not included in the fit would be present in the residual).

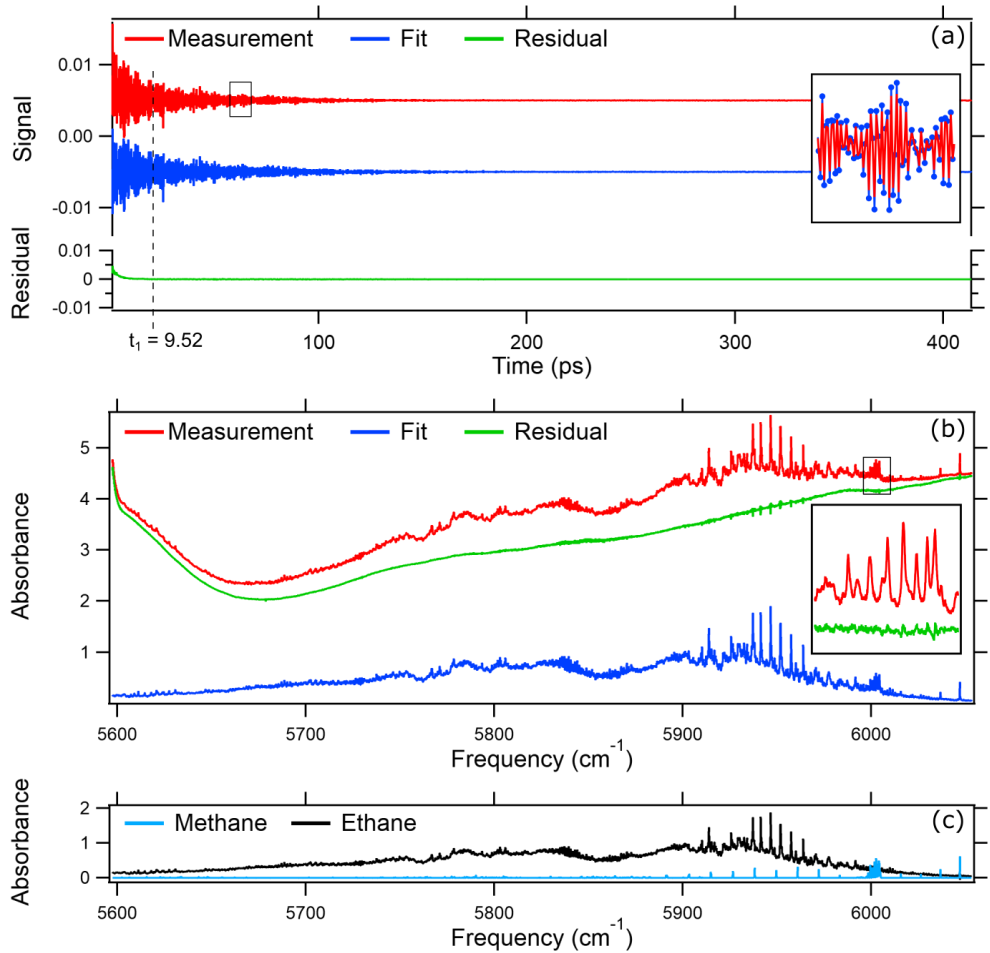

Fig. 4. The m-FID fit to the ethane-methane mixture. (a) The results from a fit to the modified free induction decay in the time-domain, where the measurement and the fitted model are offset for clarity. The chosen value for the weighting function parameter $t_1$ is indicated by the dashed line. The inset shows a direct comparison between the measured signal and the model. (b) The Fourier transform of the time-domain signals in panel (a), which are the frequency-domain representation of the fit results. The inset shows the methane Q-branch. (c) Absorbance models for ethane and methane individually, scaled to their concentrations in the known mixture.

Without any regard for the baseline source intensity, the fit converges on a mole fraction measurement of 3.4±0.1% methane and 96.6±0.1% ethane, both of which are well within the uncertainty of the known experimental mixture. The fit results are shown graphically in Fig. 5. The uncertainties in the measured mole fractions are the statistical fit uncertainty reported by the Levenberg-Marquardt algorithm combined in quadrature with the uncertainty from the pressure, temperature, and path length values that were held constant by the fitting routine. To assess this latter component of the uncertainty, we performed different fits while changing values for the pressure, temperature, and path length within their experimental uncertainties. We use the range in the ethane and methane mole fractions measured using this approach to define this component of the fit uncertainty. As shown in Fig. 5, the uncertainty due to the fixed parameters dominates compared to the statistical uncertainty of the fitting algorithm.

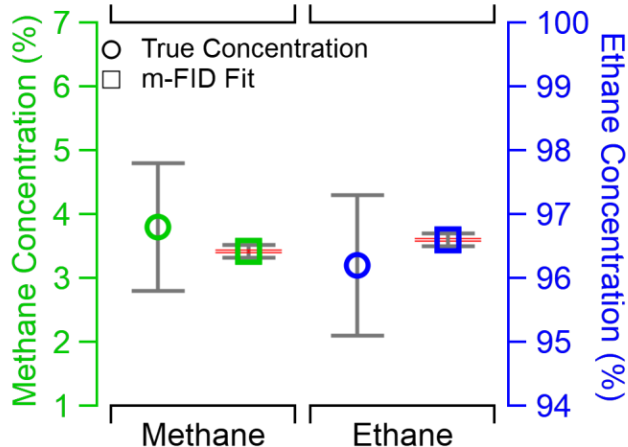

Fig. 5. Comparison of the true ethane and methane mole fractions (circles) with the mole fractions measured using the m-FID approach (squares). The uncertainty in the known mole fraction is primarily due to pressure transducer uncertainty. For the measured mole fractions, the large, gray error bars are a combination of the statistical uncertainty reported by the Levenberg-Marquardt algorithm with the uncertainty associated with the pressure, temperature, and path length that were held constant by the fitting algorithm. The thin red error bars indicate the statistical uncertainty only.

## 6. Application to complex mixtures of broadly absorbing molecules

The greatest benefit of the m-FID approach is achieved for mixtures of broadly absorbing molecules, situations with complex baselines from nonlinear broadening in lasers and window etalons, and for species with imperfect absorption models that could lead to the introduction of baseline bias in traditional frequency-domain fitting routines. To test the m-FID approach against all three challenging conditions, we simulate the complex spectrum of four broadly absorbing molecules (ethane, propane, acetone, and acetaldehyde) in the C-H stretch spectral region around 3000 cm$^{-1}$ over a 2 km path length at typical atmospheric conditions and with realistic transmission noise. The spectrum is created with published absorption cross sections [45–48], and includes strong water vapor absorption modeled with the HITRAN database [41] to simulate $H_2O$ interference common in real world measurements.

We add Gaussian noise to the modeled transmission spectrum (equivalent to approximately 0.001 absorbance noise) as well as a light source intensity profile representative of a mode-locked mid-infrared dual comb spectrometer. We include two high frequency etalons (with transmission amplitudes and free spectral ranges $A_1$=0.08, $f_1$=235 GHz and $A_2$=0.47, $f_2$=1.18 THz) to approximate intensity modulation due to nonlinear broadening processes and window interference. Figure 6(a) shows the final appearance of the simulated transmission spectrum (black trace) and the non-absorbing transmission baseline (red trace). Figure 6(b) shows the absorbance spectrum for each constituent species at the simulated conditions, and Fig. 6(c) shows the m-FID signal corresponding to each spectrum in Fig. 6(b). Figure 6(d) shows the time-domain signal corresponding to the simulated baseline intensity at short times.

To replicate real-world conditions in which model-measurement mismatch is expected, we fit the spectrum using different absorption models than those used to create the simulated transmission spectrum. In this case, we fit each simulated molecule (except water vapor) using absorption cross section data from the PNNL Infrared Database [49]. The PNNL reference spectra were measured at approximately the same temperature and pressure as the absorption cross sections [45–48] used to generate the simulated transmission spectrum. Appendix C contains a detailed comparison of the differences between the PNNL database and the cross sections used to generate the simulated spectrum. This comparison shows that the two databases differ by up to 12% for the molecules included in the simulated spectrum. While any absorption spectroscopy approach is unlikely to yield a perfect concentration retrieval when the absorption model is not correct, by eliminating the need to know or fit the baseline light source intensity, the m-FID approach removes the many degrees of freedom associated with the baseline fit in a traditional frequency-domain approach that can couple with the absorption model during the fit. We thus expect (and see in the fit results presented below) that this can improve the robustness and accuracy of the concentration retrieval under real-world conditions that include imperfect absorption models.

To demonstrate this, we also fit the simulated spectrum using a traditional frequency-domain approach. Here, it is important to note that there is no standardized method for frequency-domain baseline correction in quantitative sensing applications, and each research group tends to develop or adapt its own method best suited for the application. The frequency-domain approach we use in this demonstration involves fitting the measured absorption features with a known model while simultaneously modeling the baseline with a series of polynomials. This technique models the baseline piecewise by dividing the spectrum into sections and fitting a polynomial to the non-absorbing baseline of each section. The algorithm then simultaneously optimizes the baseline polynomial coefficients for all segments alongside the conditions of the spectral model (temperature, pressure and concentration) to deliver a measurement of the sample gas properties. To fit our simulated spectrum, we use six segments, each with a 9$^{th}$ order polynomial to account for the complex baseline. Further details on the frequency-domain approach, including a plot of the fit results, are given in Appendix C.

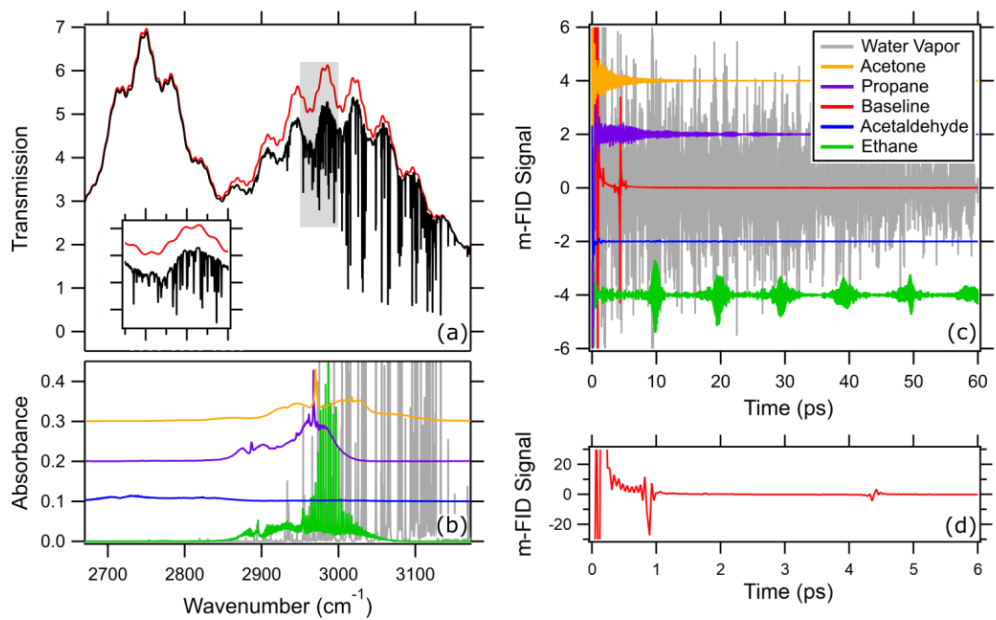

Fig. 6. Shows the simulated spectrum used to test the m-FID approach for severely overlapping, broad absorbing molecules with a non-ideal laser intensity spectrum. (a) Shows the transmission spectrum of the simulated mixture degraded with noise and a baseline laser intensity spectrum (shown in red). The inset highlights the two high frequency etalons included in the simulated baseline. (b) Shows the absorbance spectrum of each molecule in the simulated mixture assuming a path length of 2 km and the mole fractions given in Table 1. (c) Shows the m-FID signals for each spectrum in panel (b). (d) Shows an expanded view of the time-domain signal corresponding to the baseline. The sharp features near 0.9 ps and 4.5 ps are the low and high frequency etalon interference signals, respectively. Both (b) and (c) are offset for clarity.

We apply the m-FID and frequency-domain techniques to determine the absorber mole fractions from the simulated spectrum with and without the complex laser intensity baseline described above. When fitting the spectrum with the complex baseline, we set the $t_1$ parameter in the m-FID weighting function such that the two etalon signals shown in Fig. 6(d) are not included in the fit window. Table 1 summarizes the percent error in the measured mole fractions using the two techniques.

As expected, the retrieved mole fractions for the spectrum without the complex laser intensity baseline agree well between the m-FID and frequency-domain techniques. Because the spectra of the molecules overlap, the fit results for each species are coupled. This coupling allows the model errors for one species to affect the concentration retrievals for the other species. For this reason, the fit errors in Table 1 are not directly correlated to the errors between the two spectral databases used to generate and fit the simulated spectrum. Similarly, we attribute the minor differences between the no-baseline fit results using the two techniques to the fact that the molecular response signals overlap differently in the time and frequency domains. This may allow the fits to each molecule to share fit error differently, leading to minor differences in the retrieved concentrations.

After degrading the simulated transmission spectrum with a complex baseline and noise, the percent error in the mole fractions we measure using the m-FID technique remain nearly constant when compared to the fit results using the non-degraded spectrum. This indicates that the m-FID technique is robust against the effects of the added baseline and noise, even in the presence of realistic model-data mismatch (>10%), which is induced here by generating and fitting the spectrum with different absorption models. In comparison, when using the frequency-domain technique, the fit does not converge to the correct concentrations for the molecules with broad absorption spectra after adding the baseline and noise. A full version of Table 1 (including the measured mole fractions) is given in Appendix C.

The increased error using the frequency-domain approach can be attributed to three factors. First, absorption by spectrally smooth and broadly absorbing molecules (such as acetone), can be incorrectly attributed to baseline modulation by the frequency-domain fitting approach, leading to unrealistically low retrieved mole fractions of these species. Second, the simultaneous optimization of the absorption model with the baseline polynomial coefficients introduces additional degrees of freedom that must be optimized (in this case 54 polynomial coefficients). With more variables, the fit is less likely to converge on a stable solution and the baseline is more likely to couple with the molecular absorption. Finally, frequency-domain techniques can be very sensitive to measurement noise, especially when analyzing weak spectral signatures (e.g. acetaldehyde in this case). Noise can be addressed in some cases by applying a smoothing function [e.g. 21] to the measured spectrum or apodizing the signal [e.g. 50], however these techniques result in an added instrument line shape. In the m-FID approach, the noise from the frequency-domain spectrum is spread across the entire time-domain signal, of which only a small component contains the molecular absorption response. As such, the molecular response in the m-FID signal is not affected by the high-frequency noise that does not overlap with the molecular response signal. Using the m-FID approach, this noise benefit occurs without the detrimental effects of an added instrument line shape.

Table 1: Percent error in the mole fractions measured from the simulated spectrum using both the m-FID and frequency-domain techniques in both test configurations (with and without the added baseline and intensity noise). The percent error remains nearly constant after adding the simulated baseline and noise when using the m-FID approach. Errors in the measured mole fractions are much higher after adding the baseline and noise when using the frequency-domain technique.

|  |  | Acetaldehyde | Acetone | Ethane | Propane | Water |
|---|---|---|---|---|---|---|
| **Simulated Mole Fraction (ppb)** |  | 20 | 200 | 40 | 40 | 80000 |
| **No Baseline** | **m-FID** | -0.55% | 3.4% | 7.9% | 0.22% | -0.17% |
|  | **Frequency-Domain** | -2.4% | -1.9% | 7.3% | -0.10% | -0.48% |
| **Added Baseline & Noise** | **m-FID** | -2.4% | 3.4% | 7.9% | 0.30% | -0.16% |
|  | **Frequency-Domain** | -99% | -24% | 6.8% | 15% | -0.78% |

The m-FID and frequency-domain approaches also differ in terms of computational speed. When analyzing the simulated spectrum with the added baseline and noise, the m-FID algorithm converges on measured mole fractions approximately twenty times faster than the frequency-domain technique – even in a scenario where frequency-domain analysis cannot recover accurate concentrations. We attribute this large difference to the greatly decreased degrees of freedom in the m-FID fit (59 degrees of freedom for the frequency-domain approach compared with 5 degrees of freedom with the m-FID approach).

## 7. Summary and conclusion

We demonstrate a new technique for baseline-free quantitative absorption spectroscopy that can be implemented with any absorption spectrometer. The approach uses cepstral analysis of the measured transmission spectrum to generate a signal (the modified free induction decay, m-FID) in which the laser intensity is decoupled from the molecular response to the light. We use the m-FID technique to accurately measure absorber concentration from a spectrally overlapping sample of ethane and methane, which absorbs continuously for more than 500 cm$^{-1}$. We extend our results toward sensing applications using broadband, mid-infrared spectrometers by using the m-FID approach to measure concentration from the simulated spectrum of four broadly absorbing species with strong interfering water vapor absorption. In this demonstration, we show that the m-FID technique dramatically reduces error in the measured mole fractions when a complex laser intensity baseline is present, while also dramatically increasing the speed of the fitting process. The m-FID technique decreases the fit degrees of freedom compared to a traditional, frequency-domain analysis technique and reduces the need for user judgement in the fitting process.

Baseline correction is a lengthy process that, in many cases, becomes the limiting factor in the accuracy of absorption-based measurements. The m-FID approach addresses this problem by eliminating the need to correct for the spectrometer's baseline intensity spectrum. The m-FID technique can be applied to any spectrometer, and is well suited to applications seeking to measure complex, real-world mixtures involving large, broadly absorbing molecules.

## Appendix A: Estimation of the detection limit for the m-FID approach

To estimate the detection limit for the m-FID approach, we closely follow the derivation of Adler *et al.* [18], making only minor modification to apply their results to the m-FID approach. The derivation assumes an absorbance spectrum $\alpha(\nu_i)$ that is measured with uniform absorbance noise of standard deviation $\sigma_\alpha$. The absorbance spectrum is converted to the m-FID signal through a Fourier transform. The resulting m-FID signal $A(t_i)$ is sampled at times $t_i$ and contains uniform time-domain noise $\sigma_A$. The measured m-FID signal is fit with a time-domain model of the molecular response $A_o(t_i)$ that is generated at a given temperature, pressure, and absorber concentration $\chi$. The fit determines the best fit scaling factor $\beta$ between the measurement and the model, which results in the fitted absorber concentration $\beta\chi$. An estimate for the detection limit requires an estimate of the standard deviation of the scaling parameter, $\sigma_\beta$, which we derive here. The detection limit is then calculated as $\sigma_\beta\chi$.

The fitting process involves minimizing the sum of the squares of the residuals between the m-FID signal and the scaled model

$$f(\beta) = \sum_i^k W(t_i)[A(t_i) - \beta A_o(t_i)]^2 = minimum \quad (7)$$

where $W(t_i)$ is the weighting function (e.g. Eq. (6)) that is multiplied with the fit residuals to exclude the baseline contribution from the fitting routine. The minimum to Eq. (7) occurs for $df/d\beta = 0$, which gives the fitted value for $\beta$ equal to

$$\beta = \frac{\sum_i^k W(t_i) A(t_i) A_o(t_i)}{\sum_i^k W(t_i) A_o^2(t_i)} \quad (8)$$

The standard deviation of the scaling parameter $\beta$ can be estimated as

$$\sigma_\beta^2 = \sum_i^k \left[\frac{\partial \beta}{\partial A(t_i)}\right]^2 \sigma_A^2 \quad (9)$$

where

$$\frac{\partial \beta}{\partial A(t_i)} = \frac{W(t_i) A_o(t_i)}{\sum_i^k W(t_i) A_o^2(t_i)} \quad (10)$$

Combining Eqs. (9) and (10) gives the final expression for $\sigma_\beta$:

$$\sigma_\beta = \frac{\left[\sum_i^k W^2(t_i) A_o^2(t_i)\right]^{\frac{1}{2}}}{\sum_i^k W(t_i) A_o^2(t_i)} \sigma_A \quad (11)$$

In the limit that the m-FID signal is uniformly weighted at unity over the entire signal, Eq. (11) reduces to the exact form derived in [18]. Finally, the equations derived above assume a linear fit for concentration only, although the resulting equations could be adapted to estimate detection limits for multi-parameter fits [18].

We use the concentration detection limit to evaluate the effect of excluding a small portion of the molecular response in the m-FID signal from the fitting algorithm (i.e. the portion of the molecular response that is gated-out along with the baseline intensity). We calculate the detection limit using the near-IR spectrum of pure carbon monoxide simulated at 0.8 bar, 296 K, and a 1 m path length. We assume a value for the absorbance noise of 0.001. We convert the simulated absorbance noise to the corresponding time-domain noise, which we calculate as the standard deviation of the noise after applying a Fourier transform to generate the time-domain signal. We calculate the m-FID detection limit in two cases. First, we calculate the detection limit in the case that the entire molecular response can be fit (i.e. a perfectly flat baseline), which yields an m-FID detection limit of 0.02%. Second, we calculate the limit in the case that a large portion (~20%) of the m-FID signal is excluded by the weighting function, which corresponds to the case of a very highly modulated baseline. In this case, the m-FID detection limit is 0.07%. For comparison, we calculate the frequency-domain detection limit for the flat baseline scenario using the expression given in [18] to be 0.03%. We suspect that the slight improvement in the m-FID detection limit in the flat baseline case is due to the fact that while Gaussian noise distributes evenly across both the m-FID and frequency-domain signals, only a small portion of the m-FID signal contains the molecular response (as opposed to the frequency domain, where the molecular response is also spread out across the spectrum). In other words, the molecular response in the m-FID signal is not affected by the noise that does not overlap with the molecular response. As a result, the molecular response in the m-FID signal has slightly better signal-to-noise characteristics compared to the frequency-domain. However, as expected, as we increase the portion of the molecular response that is excluded by the weighting function, the information loss ultimately causes the m-FID detection limit to increase relative to the frequency-domain limit.

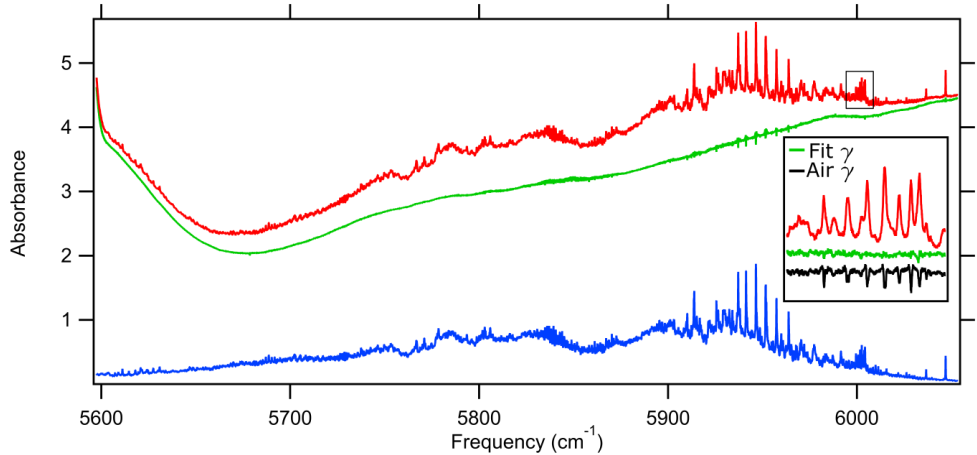

Fig 7: Shows the frequency-domain results of the m-FID fit to the ethane-methane spectrum. The inset shows the methane Q-branch. Here, the 'Air $\gamma$' trace is the residual when air-broadening of methane lines is used. The 'Fit $\gamma$' trace is the residual with the scaled methane-broadening coefficients.

**Appendix B: Spectral model to fit the ethane-methane spectrum**

We model the methane absorption features in the ethane-methane mixture spectrum using the HITRAN 2016 database [41] with a Voigt profile generated using the HAPI interface [44]. We include the first and second isotopologues of methane ($^{12}CH_4$ and $^{13}CH_4$) due to the relatively high natural abundance of $^{13}CH_4$ (>1%) [41]. Broadening of methane absorption features by ethane is not published in spectral databases, and assuming air-broadening for the methane lines leads to an observed increase in the fit residuals (see black trace in the inset to Fig. 7). To approximate ethane broadening, we fit a band-wide scaling factor to the methane air-broadening coefficients to account for broadening of the methane lines by ethane. The inset to Fig. 7 shows that the scaled broadening coefficients significantly decrease the residual in the fit to the methane Q-branch.

Line shape parameters for ethane are not available for most lines in the near-infrared, and the ethane absorption coefficients published in the PNNL Infrared Database [49] are at 0.112 cm$^{-1}$ spectral resolution (significantly lower than the 0.0066 cm$^{-1}$ point spacing of the dual-comb measurement shown in this paper). We instead model the ethane component of the methane-ethane mixture using an air-broadened ethane spectrum measured at room temperature and 612.5 Torr with a high-resolution dual-comb spectrometer at NIST Boulder (unpublished).

**Appendix C: Further details of the concentration fits to the simulated mid-infrared spectrum**

*Comparison of absorption models used to simulate and fit the spectrum*

To create the simulated mid-infrared spectrum of four overlapping and broadly absorbing compounds (ethane, propane, acetone, and acetaldehyde), we use previously published, high-resolution absorption cross section data from references[45–48]. We fit the simulated spectrum using reference spectra from the PNNL Infrared Database [49]. To match the resolution of the two data sources for the fits, the high resolution cross section data of [45–48] is resampled to match the point spacing of the PNNL Infrared Database reference spectra, which were measured at 0.112 cm$^{-1}$ spectral resolution. By generating and fitting the simulated spectrum with different absorption models, we introduce realistic model-data mismatch that is expected for real world measurements, where the absorption model is not a perfect match to the actual spectrum that is measured by the spectrometer. While the two absorption models were published at similar temperatures and pressures, differences upwards of 10% remain between the two models which are likely due to a different ratio of air-broadening to self-broadening conditions between the datasets. Figure 8 shows the differences between the two absorption models, which vary by up to 12%.

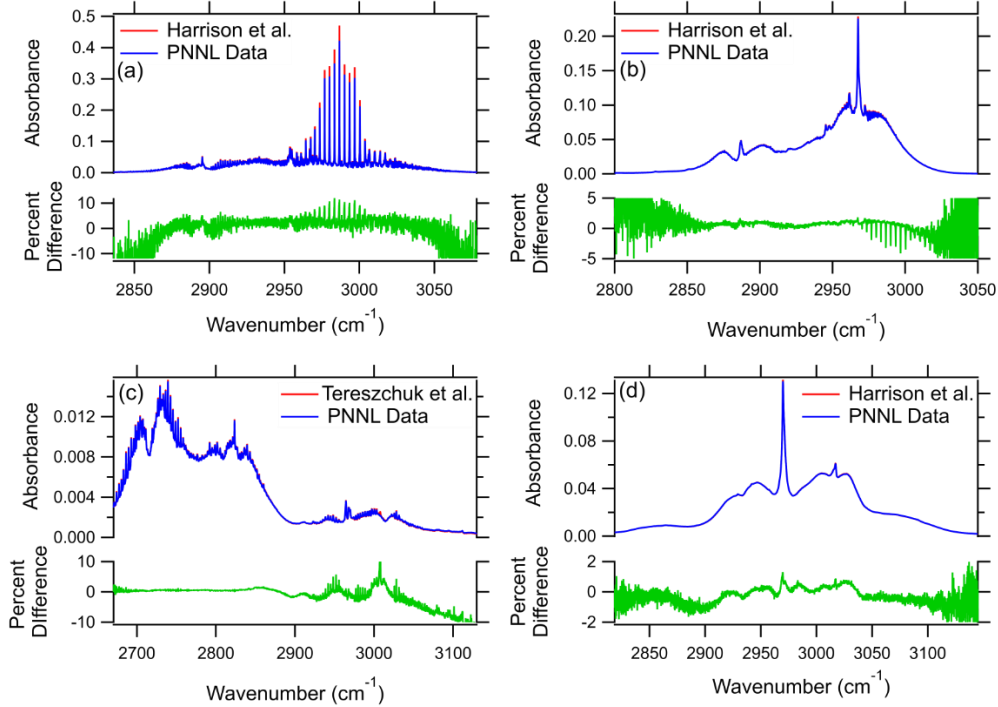

Fig. 8: Comparison between the two spectral models used to create and fit the simulated mid-IR measurement. In each panel, the percent difference is evaluated relative to the PNNL Infrared Database. (a) Ethane spectrum from Harrison et al. (2010) [45] compared to the PNNL Infrared Database. (b) Propane spectrum from Harrison et al. (2010) [46]. (c) Acetaldehyde spectrum from Tereszchuk et al. (2011) [48]. (d) Acetone spectrum from Harrison et al. (2011) [47].

*Frequency-domain fitting approach and comparison of fit results*

For comparison to the m-FID approach, we also fit the simulated MIR spectrum using a traditional frequency-domain approach. As mentioned in Section 6, the frequency-domain fitting algorithm fits the spectrum with a known, condition-specific absorption model while simultaneously modeling the baseline using a series of polynomials. Specifically, the approach divides the spectrum into sections, and then fits each section with a polynomial to account for the non-absorbing baseline. A Python-based Levenberg-Marquardt algorithm optimizes the baseline polynomial coefficients for all sections along with the parameters of the spectral model to deliver a measurement of the absorber temperature, pressure, or concentration.

We simulated the complex baseline shown in Fig. 6 by adding and multiplying several functions (sinusoids, polynomials and Gaussians). We then added two sine waves of different frequencies to generate two etalon interference signals. In order to fit this complex baseline using the frequency-domain approach, we used a set of six, $9^{th}$ order polynomials. The number and degree of the polynomials were chosen by trial and error to minimize the baseline contribution in the fit residuals. We attempted to include sine components to the baseline fit to account for the high frequency etalon, but despite significant effort were not able to achieve a stable fit. Even when constraining the etalon amplitudes and frequencies to reasonable ranges around their known values, the fitting algorithm returned unrealistic values for the etalon amplitudes. We suspect that the added degrees of freedom due to the etalon components in the baseline model (amplitude, frequency, and phase) prevent a reliable fit when an etalon is included in the baseline model.

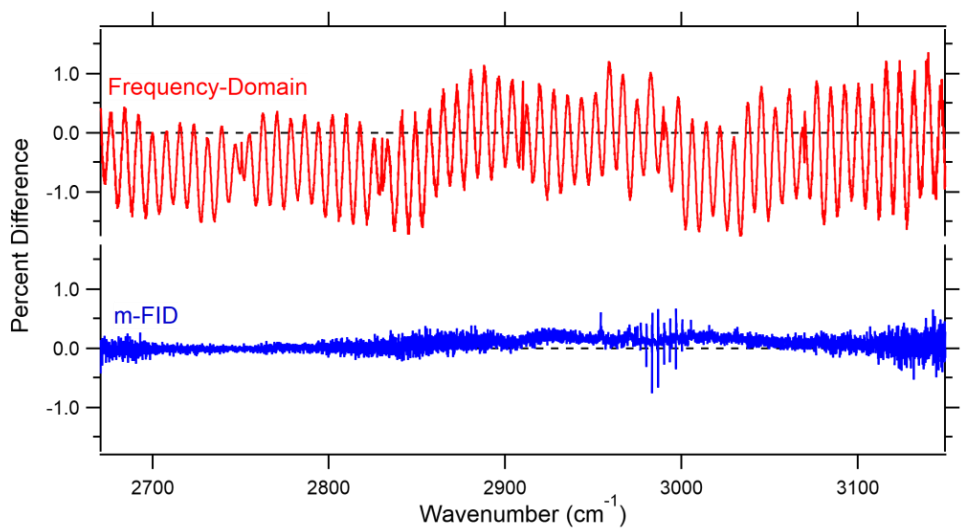

Fig 9: Red trace: the percent difference between the baseline determined by the frequency-domain algorithm and the simulated baseline. Blue trace: the percent difference between the m-FID residuals (containing both the baseline and the mismatch between the measured and modeled absorption features) and the simulated baseline. The frequency-domain comparison shows that the high frequency etalon is not removed, and the deviations away from zero indicate remaining baseline contribution that is not fully modeled. The m-FID comparison does not indicate significant remaining baseline contribution, and also shows remaining differences between the measured and modeled absorption features.

To assess the accuracy of the frequency-domain baseline model, we calculate the percent difference between the baseline determined by the frequency-domain algorithm and the true, simulated baseline. This comparison is shown in the red trace in Fig. 9. As discussed above, Fig. 9 clearly shows that the high frequency etalon has not been removed, however the $9^{th}$ order polynomials have largely compensated for the lower frequency etalon. Additionally, we see that there are still significant deviations away from zero, which indicate that there is remaining baseline contribution that is not fully removed by the frequency-domain baseline model. As mentioned above, the choice of six, $9^{th}$ order polynomials was found to minimize this remaining baseline contribution.

To compare the frequency-domain fit results to the m-FID approach, we also calculate the percent difference between the Fourier transform m-FID fit residuals and the simulated baseline. Here, it is important to note that the m-FID fit results contain *both* the baseline intensity and the remaining differences between the measured and modeled absorption features. Furthermore, we were able to exclude the etalon effects from the m-FID fitting algorithm by setting the $t_1$ parameter in the m-FID weighting function (Eq. (6)) such that the two etalon signals were not included in the m-FID fit. The m-FID results are shown as the blue trace in Fig. 9. The m-FID results are largely flat and centered at zero, which is indicative of no significant remaining baseline contribution. The remaining discrepancies are due to the differences between the measured and modeled absorption features.

*Complete Table of Fit Results*

Table 2 gives the complete fit results for the simulated, mid-infrared spectrum showing the concentrations measured using both the m-FID and frequency-domain analysis techniques.

Table 2: Shows complete fit results (mole fractions and errors) from the simulated spectrum of acetaldehyde, acetone, ethane, propane, and water vapor.

|  |  |  | Acetaldehyde | Acetone | Ethane | Propane | Water |
|---|---|---|---|---|---|---|---|
|  |  | **Simulated Concentration (ppb)** | 20 | 200 | 40 | 40 | 80000 |
|  |  |  |  |  |  |  |  |
| **No Baseline** | **m-FID** | Conc. (ppb) | 19.9 | 207 | 43.1 | 40.1 | 79900 |
|  |  | % Error | -0.55% | 3.4% | 7.9% | 0.22% | -0.17% |
|  | **Frequency-Domain** | Conc. (ppb) | 19.5 | 196 | 42.9 | 40.0 | 79600 |
|  |  | % Error | -2.4% | -1.9% | 7.3% | -0.10% | -0.48% |
|  |  |  |  |  |  |  |  |
| **Added Baseline & Noise** | **m-FID** | Conc. (ppb) | 19.5 | 207 | 43.1 | 40.1 | 79900 |
|  |  | % Error | -2.4% | 3.4% | 7.9% | 0.30% | -0.16% |
|  | **Frequency-Domain** | Conc. (ppb) | 0.126 | 152 | 42.7 | 45.8 | 79400 |
|  |  | % Error | -99% | -24% | 6.8% | 15% | -0.78% |

## Funding

This work was funded by the Air Force Office of Scientific Research (FA9550-17-1-0224), the Defense Advanced Research Projects Agency (W31P4Q-15-1-0011 from AMRDEC), the Strategic Environmental Research and Development Program (W912HQ-16-C-0026), and NASA Headquarters under the NASA Earth and Space Science Fellowship program (PLANET18R-0018).

## Acknowledgments

The authors wish to thank Dr. Eleanor Waxman for providing the ethane reference spectrum, and Drs. Ian Coddington and Fabrizio Giorgetta for helpful discussions. We acknowledge the insightful comments of our anonymous reviewers.

## Disclosures

The authors declare no conflicts of interest.